# A Driven Backup Routing Table to Find Alternative Disjoint Path in Ad Hoc Wireless

Radwan S. Abujassar [1], Mohammed Ghanbari[2], Fellow, IEEE

[1]School of Computer Science and Electronic Engineering, University of Essex, Essex, Colchester, UK

rabuja@essex.ac.uk

[2]Essex, Wivenhoe Park, CO43SQ

ghan@essex.ac.uk

**ABSTRACT.**

*The performances of the routing protocols are important since they compute the primary path between source and destination. In addition, routing protocols need to detect failure within a short period of time when nodes move to start updating the routing table in order to find a new primary path to the destination. Meantime, loss of packets and end-to- end delays will increase thereby reducing throughput and degrading the performance of the network. This paper proposes a new algorithm, DBRT (Driven Backup Routing Table), to improve the existing proactive protocols such as DSDV (Destination Sequenced Distance Vector) protocol by creating a backup routing table to provide multiple alternative routes.  The DBRT algorithm identifies adjacent nodes for each node in the same range and then selects one of these as a backup next hop according to the available path to the destination. The results show that loss of data packets, throughput and end-to-end delay times between source and destination are improved. The results show that the new protocol does not degrade the network's performance despite sending extra messages to construct and update the new backup routing table. Simulations (using an NS2 simulator) are undertaken to demonstrate the difference between using a DSDV protocol with or without the proposed schema.*

*Index Terms; Network Protocols ;( DSDV) Destination Sequenced Distance Vector ; DBRT (Driven Backup Routing Table);Wireless Network; Mobile Ad Hoc Network*

## 1 INTRODUCTION

In Mobile Ad Hoc Networks (MANET) is unlike the wired networks, because there is no central infrastructure between the nodes. Each node can exchange data dynamically without the need to a fixed base station or a wired back- bone. Some limitations of the MANET network have been researched, such as, transmission power limitation and multiple hops. This is because MANET
Uses intermediate nodes to exchange information to pass its traffic to its destination. Hence, route discovery and maintenance in MANET networks is an essential issue. The nodes in wireless ad hoc networks can move frequently and instantaneously from area to area without notification, which leads to various problems, such as, loss of connectivity and an increase in the holding time, during which a new shortest path between source and destination is computed for the routing table[1]. When a loss in connectivity occurs, not all the nodes on the topology will be informed. This will generate loops in the network, which degrade its performance and reduce throughput. IP recovery will discover a backup path within a short period to alleviate loss of packets, reduce end-to- end delay and avoid loop in the network [3]. In MANET ad hoc networks, there are various kinds of main routing protocol tables. In table-based proto- cols, each node constructs a routing table that includes all routes to all nodes on the topology. The routing protocol needs to send periodic messages that contain routing information to keep the routing table for each node up to date. In on demand-protocols, nodes compute routes when they are needed. Ad hoc wireless networks are frequently affected by failures when nodes move in and out and of radio propagation range. It is, therefore, highly desirable to develop a recovery mechanism to improve the quality of service (QoS) of the network. In the meantime, loss of data packets and end-to-end delays will increase. Many different types of routing protocols have been used to solve this routing problem, including DSDV, Dynamic Source Routing (DSR) and Optimized Link State Routing (OLSR) protocols [4]. In wired networks, the routing protocol generally uses distance vectors or link state routing algorithms. Both are proactive mechanisms as they send extra messages to keep the nodes up-to-date in case any information on the network changes, such as, if a node joins the network or it fails. When failure occurs, these protocols inform all the nodes and they start to re-compute a new routing table. More holding time is then required in order to re-send the traffic along the new route. In this paper, we propose a new algorithm called Driven Backup Routing Table (DBRT) to improve the existing proactive routing protocols, such as, DSDV and OLSR to construct a backup routing table based upon its original that consists of a backup path for each node to its destination on the topology. When nodes fail or changed their positions by moving out of range, DSDV and LS (Link State) protocols demand that a routing advertisement be broadcast between the nodes on the network. In DSDV, when the nodes receive these advertisements, each one knows the route from its neighbor and its distance to all the other nodes on the network. On the other hand, OLSR protocols compute the shortest path based on the complete picture for each adjacent node on the network. The DBRT mechanism aims to recover the network from failure in a shorter period by pre-computing a backup routing table by considering more than one node that have moved or change their positions. The backup routing table has alternative paths along which to pass the traffic when failure occurs. The pre-computed alternative path can be used immediately without waiting for the routing protocol to re-compute a new one. In this paper, we concentrate on the scenario when more than one node moves on the primary path. The main contribution of this paper is towards the development of an alternative and fully disjointed pathway, which is computed. By using a backup routing table and is based upon the number of adjacent nodes and their ranges.

This paper is organized as follows: Section 2 discusses related work, Section 3 illustrates the originality and the basic concept for the DBRT algorithm in detail, Section 4 shows the performance evaluation via simulation and concluding remarks are given section 5.

## 2 RELATED WORK

Several routing protocols have been published for use in different environments to improve network performance when nodes move or fail, causing a loss in connectivity [7]. In [8] the author studies the QoS and reliability in the MANET networks. There are two categories of routing protocols for MANETs namely proactive and reactive. A proactive routing protocol updates pairs of nodes by flooding via a periodic broadcast. This brings routing tables up-to-date for each node in the network. However, a reactive routing protocol detects a new route only when it is required. Some proactive routing protocols (such as DSDV, OLSR, CGSR and WRP) trigger messages that can detect links when they fail [9, 10]. Based on these messages, the routing protocol can construct and maintain routes to the destination. Reactive protocols, such as, DSR, AODV (On-demand Distance Vector) and TORA will reduce overhead because new paths between nodes will be created only when failures occur. The DSDV routing protocol is based on the Bellman-Ford algorithm. It is similar to a DSR protocol, which is an AODV protocol [11]. This is because they both use a similar algorithm. Multipath routing has been introduced and considered in wireless ad-hoc networks for improving QoS and reliability of network performance [12]. In [13] we present a new algorithm for wired networks called the Alternative Routing Table (ART) algorithm, which sends messages to enquire if neighboring nodes have an alternative and disjointed pathway from the primary node to its destination. The number of data packets sent to each node will depend on the number of adjacent nodes that are not connected to the primary path. An efficient routing protocol algorithm has been constructed in terms of achieving robustness and fast convergence in case a node goes down. As such, we have enhanced the ART algorithm to make it work on wireless ad-hoc networks by constructing a primary routing table based on the propagation range with the number of hops and a backup routing table based upon the number of adjacent nodes to select the best path. In [14–16], the OLSR routing protocol is shown to be proactive in ad-hoc networks. The OLSR has Multi Point Relay (MPR) nodes, which are used to send link state messages to construct a routing table [17]. In OLSR, two kinds of broadcasts are sent: HELLO and Topology Control (TC) messages. Each node will send HELLO message to its neighbours to check if connectivity is up or down every two seconds but its waiting time of six seconds is considered too long. The TC message is thus based on the information collected by the HELLO messages. The interval time is five seconds and the time to detect failure is fifteen seconds.

In [18, 19], the source node knows the complete rout hop-by-hop to the destination. The route from source to destination is stored in a route cache. The OLSR routing protocol will re-route data packets only after re-computing a new routing table and updating the information to all nodes on the topology. Depending upon the HELLO message interval times, the re-routed traffic will take longer and this will lead to an increased loss of data packets and reduced throughput [20]. Internet Protocol recovery emphasizes two cases. First, the time required to detect failure and secondly, the time taken to compute the shortest path. In [21], the authors mention how a network recovery can be achieved within a short time when failure occurs. The aim of IP recovery is to offer a loop free protection mechanism in the network. Loops in a network are one of the main problems with some existing techniques. The constraint-based routing protocols use metrics instead of the shortest path between nodes to find a suitable route. In QoS routing schema, the Core Extraction Distributed Ad Hoc Routing (CEDAR) algorithm has been introduced for a medium size ad-hoc network. The idea behind the multiple paths QoS routing schema is to try to find a number of paths between source and destination based on high capacity bandwidth requirements. Service providers guarantee a network's performance via a set of measurements, such as, delay, jitter and loss of data packets [22, 23]. These parameters are part of QoS that can optimize along the network to invest in the provisioning of resources in cases of increased traffic or node failures.

## 3 PROPOSITION

The basic principle of the DBRT algorithm is how to construct a backup routing table by computing an alternative path to the destination by investing the original routing table, which is computed by the routing protocol. Our proposition leads to reduced holding times for source nodes until the new routing table is updated. The backup routing table is restricted because it depends on the adjacent nodes having a disjoint path to their destination. This gives each node the ability to offer its path for its traffic reaches its destination. However, the backup path is pre-computed in advance in order to re-route data packets in case of failure. In addition, when a node on the primary path goes down or moves from the area, the adjacent nodes will detect a link break by receiving a link layer feedback signal from the HELLO data packets, which will confirm the failure because it does not receive any acknowledgement during the interval time. The DBRT algorithm, via the backup routing table, will pass traffic to destination D without waiting to re-compute a new routing table. However, we assume here that each node has at least one adjacent one unconnected to the primary path but within the same range. The DBRT algorithm involves choosing one of them as a backup node to re-route a packet through it when a failure occurs. We performed a different number of topologies each having a different number
 Of nodes. In each topology we made each node to have at least two neighbours that can re-route data packets through them when failure occurs.

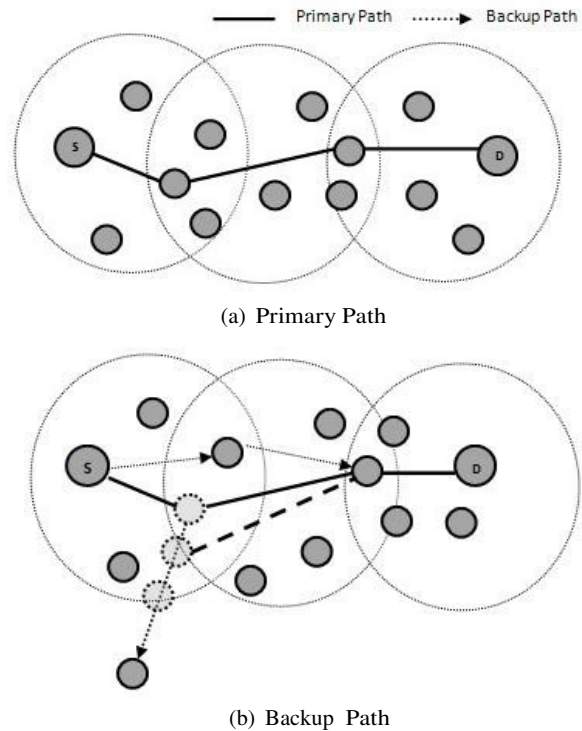

(a) Primary Path

(b) Backup Path

Fig. 1: Primary and Backup Path

### 3.1 ALGORITHMS OVERVIEW

In our schema, the pre-computed second paths provide a good solution for when failures occur. DBRT evaluates the backup routing table from the primary one to ensure that any alternative path will be disjointed from the primary one. Figure 1 shows the primary and backup path. When the routing protocol determines the primary path the DBRT algorithm will start to check each node to determine how many adjacent nodes are within range and not connected to the primary path. The DBRT computes the backup routing table from the passing traffic. In fig 1 (a), the primary path from source to destination is computed by a routing protocol. By using the radio propagation range the DBRT can discover any adjacent nodes by which to compute the backup path to the destination, which is based upon the number of hops. When there is more than one adjacent node, the DBRT will start to check which adjacent node has a disjointed path to the destination (i.e. to construct a backup path). In addition, the DBRT algorithm evaluates an appropriate backup route using the primary routing table. Any nodes connected to the primary path will be excluded from the next hop in the backup route. Each node randomly takes a position (X, Y) in the radio propagation range for the area 1000m by 800m. In DBRT, the nodes on the topology will start broadcasting a small message to enquire from adjacent nodes if they have a disjointed path from the primary (with regard to the primary routing table) to the destination. Because DBRT considers all nodes as a source and destination

49

then each node will check the acknowledgements from adjacent nodes to see if there is any available route to the destination and if one of the adjacent nodes become the intended node it will then insert it into the backup routing table as a first hop. If no adjacent node has a disjointed path, then each adjacent node will check other nodes from its list of adjacent nodes to determine if they have a neighboring node that has a disjointed route to the destination. Hence, the adjacent node will start checking its route hop-by-hop to the destination.

If this route is not connected to any node on the primary path then it will inform the source whether a disjointed path is available. Hereafter, the source node uses the node in the backup routing table, as a first next hop to re-route the traffic through the backup routing table should a node on the primary path fail. If more than one alternative path exists, then the node will select the best one based upon the number of hops that have a greater history for reliability.

The process at each node on the topology can be described as follows:

**Step 1** Each node inserts her adjacent in the adjacent list from the primary routing table but excluding nodes that are connected in the primary path.

**Step 2** Each node broadcasting to her adjacent nodes a mini packets to en- quire if they have a unique route to destination not overlap with the primary one.

**Step 3** When all adjacent nodes received these packets then they will start to check from their routing table if their route to destination not connected with primary path. If"Yes" then will insert all these next hops to the backup routing table.

**Step 4** If the adjacent nodes do not have any disconnected route then they will start checking their neighbours to see if they have available path or not. The DBRT will return to step 1 until all nodes make a full view for the topology and created the backup routing table.

## 4 SIMULATION EXPERIMENT

### 4.1 SIMULATION ENVIRONMENT

Network simulation (NS2) was performed to evaluate the proposed algorithm. We compared the simulation results of the DSDV protocol with and without our schema. NS2 offers good support for node mobility in ad-hoc networks. Different network scenarios were created, each having a different number of mobile nodes to demonstrate the effect of node movement or failure during simulation time.
We configured an area of 1000 m by 800 m. Nodes could move randomly by any
node on the primary path. A radio propagation range with transmission power of 0.28 watt was used, allowing each node to send or receive a packet to or from its neighbours for a distance of up to 250 m. For each scenario, the simulation time ran for 250 seconds.

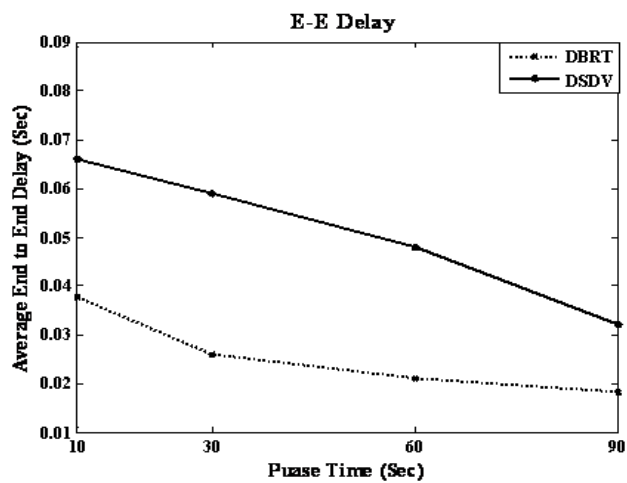

Fig. 2: End to End Delay

In the free space model, the signal power is weakness by a factor $1/d^2$ where d is the distance between radios. The movement type was of a Two Way Ground Model with a channel capacity of 2Mb/s. The packet size was 512 bytes. We compared DBRT with DSDV because they are both proactive protocols. We used an IEEE 802.11 Distributed Coordination Function (DCF), as the wireless channel can share in an ad-hoc configuration.

### 4.2 RESULTS AND ANALYSIS

Figure 2 shows the results for the end-to-end delay measured against a different number of mobile nodes. The end-to-end delay in DSDV with a DBRT algorithm was less than the DSDV protocol, because the latter has to retain all its routes in a static routing table. When any node on the primary path moved out of range or failed, the DSDV routing protocol needed to re-compute a new routing table by flooding packets to update it. In DBRT, the backup path is pre-computed in case any nodes on the primary path moves or fails. In this case, the node that is connected to the newly failed or moved node will re-route the traffic according to the backup routing table. Figure 3 shows the amount of the congestion in the network. Congestion in DSDV is less compared to DSDV with a DBRT algorithm, because DBRT needs to generate extra data packets so that it can construct a new backup routing table. These packets show that an increase in network overheads does not degrade network performance with respect to the number of nodes that are involved.

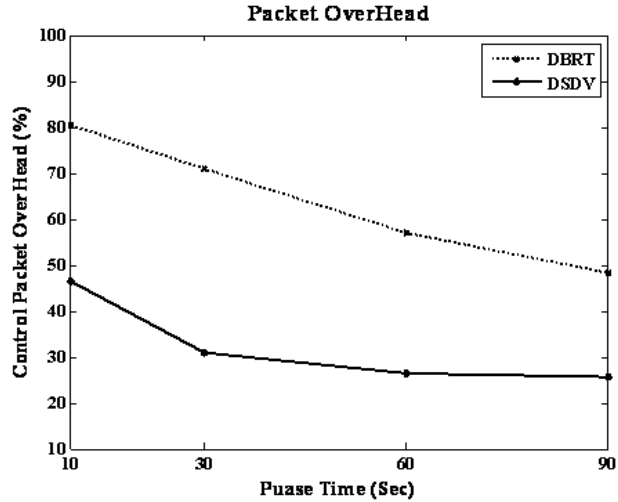

Fig. 3: Packet Overhead

Figure 4 shows the average loss of data packets for DSDV and DBRT for a different number of nodes in different scenarios. Traffic is re-routed along an Alternative path, which is computed by the DBRT protocol. The DSDV protocol shows that the loss of data packets increases based upon the number of nodes and hops to the destination. The DSDV protocol generates messages that maintain the routes that offer the greatest probability for collision to occur in the network.

Figure 5 shows the throughput for both DSDV with a DBRT algorithm and DSDV protocols. The packets will send from source to destination hop-by-hop. The DSDV protocol with a DBRT algorithm achieved a higher throughput compared to the DSDV. When a collision occurred, through- put was decreased in both. The IEEE 802.11 sending RTS packets can reduce collisions in the network. However, the DSDV protocol needs a certain amount of time to re-route the traffic. The length of this period remains undesirable. It takes two seconds for each node to re-compute a new routing table and a medium topology will take fifteen seconds.

The backup routing table will start to be calculated after the primary one has been computed. Therefore, DBRT will re-route the traffic directly to an adjacent node on the backup routing table if any node on the primary path moves out of range or fails. This will lead to an increased throughput between the source and destination.

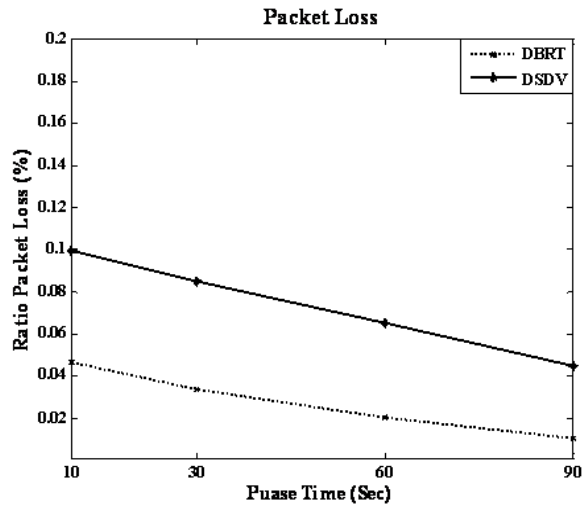

Fig. 4: Loss Of packets

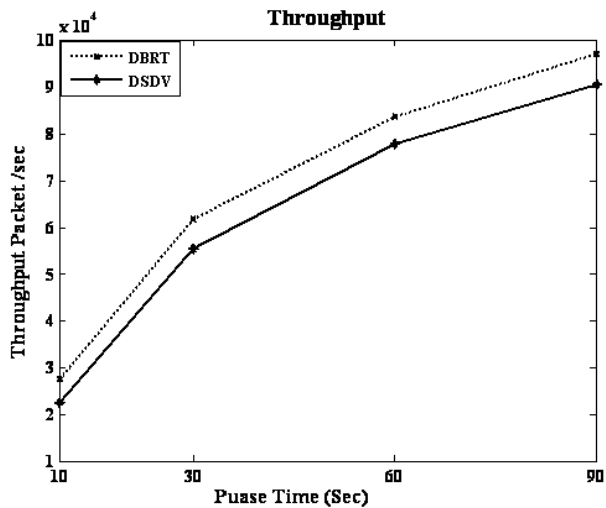

Fig. 5: Throughput

Figure 6 shows average traffic load versus pause time, (10, 30, 60 and 90s) of a 50 node network. The degree of traffic load using the DSDV protocol shows that it is satiable for their network topologies regarding the amount of packets send to update the routing table when nodes move. On the other hand, DSDV with a DBRT algorithm shows that the load is reduced when the nodes become more stable as the pause time become larger. If the pause time is 10 sec, the DBRT needs to send extra packets every 10 sec to re-update the backup routing table until the node stops. During this process, the DSDV with a DBRT will pass the traffic through an alternative path to its destination. Hence, the traffic load will increase when nodes move frequently and within a short time and vice verse.

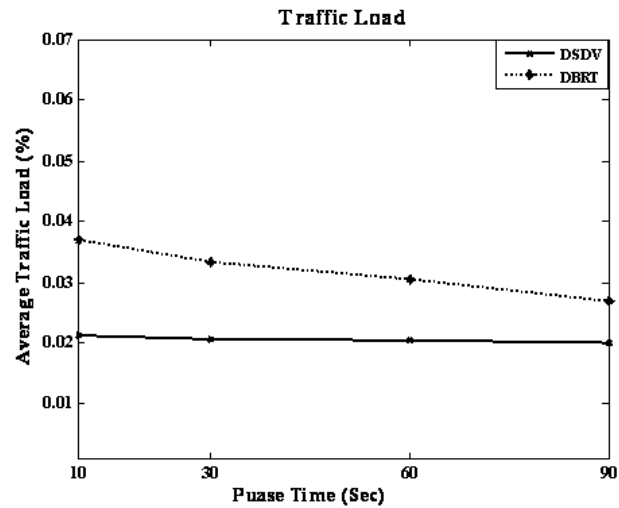

Fig. 6: Traffic Load

## 5 CONCLUSIONS

This paper has presented a new protocol for computing an alternative backup routing table which finds an alternative path for each node on the network. The DBRT algorithm computes a backup routing table based on the distance between nodes and number of hops to the destination. The backup routing table includes alternative and disjointed next hop to destination for all nodes on the topology. The DSDV protocol with DBRT provides a recovery path that gives the shortest distance between source and destination. We have shown that the backup paths contain fewer numbers of hops compared with those produced by the DSDV protocol. For real traffic, the results show that the DSDV protocol with DBRT reduces the loss of data packets and delay between source and destination nodes. In future work, the DBRT algorithm will be designed to deliver a backup routing table that contains more than one backup path between the source and destination, in order to improve the QoS when more than one node moves or fails. In addition, the reducing of sending extra packets will increase the reliability of our algorithm through creating a super node that can make updating with reduce flooding in the network.